\documentclass{rsifpublic}
\usepackage{graphicx}
\usepackage{epstopdf}
\begin{document}

\runningheads{Ricard  V.    Sol\'e et al.}{Technological singularities}

\begin{topmatter}{
\title{On singularities in combination-driven models of technological innovation}

\author{Ricard V. Sol\'e$^{1,2,3}$\corrauth, Daniel R. Amor$^{1,2}$ and Sergi Valverde$^{1,2}$\corrauth}

\address{(1) ICREA-Complex Systems Lab, Universitat Pompeu Fabra, Parc
  de Recerca Biomedica de  Barcelona. Dr Aiguader 80, 08003 Barcelona, Catalonia, Spain\\ 
  (2) Institute of Evolutionary Biology, UPF-CSIC, PsM Barceloneta 37, 08003 Barcelona\\
(3) Santa Fe Institute, 1399 Hyde Park Road, 87501 Santa Fe, New Mexico, USA}

\begin{abstract}
It has been suggested that innovations occur mainly 
by combination: the more inventions accumulate, the higher the probability that new 
inventions are obtained from previous designs. Additionally, it has been conjectured that 
the combinatorial nature of innovations naturally leads to a singularity: at some finite time, the number of innovations should 
diverge. Although these ideas are certainly appealing, no general models have been yet developed to 
test the conditions under which combinatorial technology should become explosive. Here we 
present a generalised model of technological evolution that takes into account two major properties: the number of 
previous technologies needed to create a novel one and how rapidly technology ages. 
Two different models of combinatorial growth are considered, involving different 
forms of ageing. When long-range memory is used and thus old inventions are available 
for novel innovations, singularities can emerge under some conditions with two phases 
separated by a critical boundary. If the ageing has a characteristic time scale, it is shown 
that no singularities will be observed. Instead, a "black hole" of old innovations appears and 
expands in time, making the rate of invention creation slow down into a linear regime. 
\end{abstract}


}\end{topmatter}

\corraddr{ ricard.sole@upf.edu}

%%%%%%%%%%%%%%%%%%%%%%%%%%%%%%%%%%%%%%%%%%%%

\section{Introduction}

%%%%%%%%%%%%%%%%%%%%%%%%%%%%%%%%%%%%%%%%%%%%

Technology is one of the most obvious outcomes of human culture. Over the 
last 100.000 years, humans have been able to manipulate 
their environments and the species they used to interact with in 
an amazing range of ways. Technological 
inventions have been developing at an accelerated rate since the industrial revolution [1-5] and 
economist Brian Arthur conjectured that such rapid growth is a consequence of the underlying 
dynamics of combination that drives the process [2]. Specifically, it has been suggested 
that novelties arise mainly as a consequence of new forms of interaction between 
previous artifacts or inventions [2]. Such view connects the pace of man-made evolutionary designs 
with a basic principle of biological evolution: the presence of tinkering [6] as a dominant 
way of generating new structures [7,8].

\begin{figure}[htbp]
\begin{center}
\includegraphics[width=0.48\textwidth]{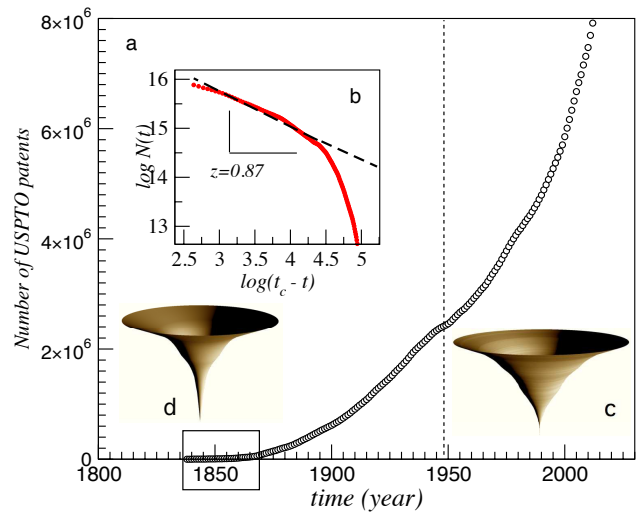}
\caption{Hyperbolic growth of the total USPTO dataset (a). 
In (b) we plot of the logarithm of patent counts against $t_c-t$ (b) for $t>1950$. The best fit 
for this data set gave an estimate $\eta=0.87 \pm 0.06$ consistent with an 
exponent $z \approx 1.14$. The inset (c) provides a picture of the spindle diagram for $N(t)$ and 
in (d) we show the same diagram for the initial part of the time series. }
\end{center}
\end{figure}

Systematic studies of technological change are difficult to perform due to 
a number of problems. These include the lack of a genome-like description of artifacts 
and the complex nature of their design paths. 
The study (largely naturalistic) of some particular systems, such as cornets [9-10] reveals 
some interesting similarities, while uncovering deep differences with cultural change.  More recent 
work based on network theory [11] provided a novel quantitative approach to technological change that defines 
a formal framework to explore technological change and the impact of design principles. 

 One consequence of the combination principle proposed by Arthur is that the growth dynamics of 
 inventions would be faster than exponential (or super-Malthusian) and should 
 exhibit a finite-time singularity [12]. The implications of such rapidly accelerating innovation processes 
 have been discussed in recent years, raising controversial speculations [13]. Predicting the progress 
 of technological change is a timely issue but also a difficult task. Nevertheless, some insights have 
 been gathered from using proper databases and statistical methods [14,15]. 
 
 A surrogate of the ways in which innovations take place in 
time is provided by patent files [16-18]. Patents are well-defined objects
 introducing a novel design, method, or solution for a given problem or set of problems. Existing data 
bases store multiple levels of patent description and they can be analyzed in full detail. 
Additionally, they indicate what previous novelties have been required to build new ones. An example is given by the U.S. Patent and Trademark Office (USPTO) patents filed 
between 1835 and 2010 \footnote{http://patft.uspto.gov}. In figure 1a we display the total number $N$ of filed  patents, 
which clearly reveals an accelerated trend over time [3,12]. The dashed line in particular indicates the start of the modern 
information technology era (around 1950). 

In [18] a study of the USPTO database was made in search for evidence of 
combinatorial evolution. The authors concluded that truly new technological capabilities 
are slowing down in their rate of appearance, but nevertheless a great deal of 
combination is present thus allowing for a "practically infinite space of technological 
configurations" [18]. One potential outcome of this virtually exploding space is a 
growth dynamics displaying potential singularities, i. e. divergent numbers of inventions 
would eventually occur as we approach a finite time window. In this paper we want to address the problem of how to define the conditions for 
technological singularities to be expected. Two main components of combinatorial dynamics will 
be taken into account: (a) the diversity (number) of potential innovations required to 
trigger a new one and (b) the degree of ageing that makes older innovations less likely to be 
used. As shown below, two main phases are expected in this diversity-aging space, 
defining the conditions for singularities to be present.

\section{Hyperbolic dynamics: minimal model}

Using the simplest approximation, we assume a neutral model of innovation based on
 pairwise combinations of existing designs. In this model,  the set ${\bf \pi} = \{ \pi_1, \pi_2, ..., \pi_{n(t)} \}$ 
defines the "design universe" at any  given time $t$.  Each $\pi_j \in \bf \pi$ 
represents an invention as described, for example, by a patent file. Here, $N(t)$ is the total number of 
inventions (patents) at year $t$. Two given designs will then combine at a given time $t$  with a given probability: 
$\pi_i + \pi_j  \buildrel  \mu_{ij} \over \longrightarrow \pi_{n(t)+1}$ 
where $ \mu_{ij}$ weights the likelihood of such an event to happen. 
This defines a second-order (bimolecular) reaction kinetics [19-21]. Such nonlinear reaction dynamics 
seems to pervade the super-exponential growth observed in a number of economic and demographic systems 
 [22-24].

\begin{figure}[htbp]
\begin{center}
\includegraphics[width=0.48\textwidth]{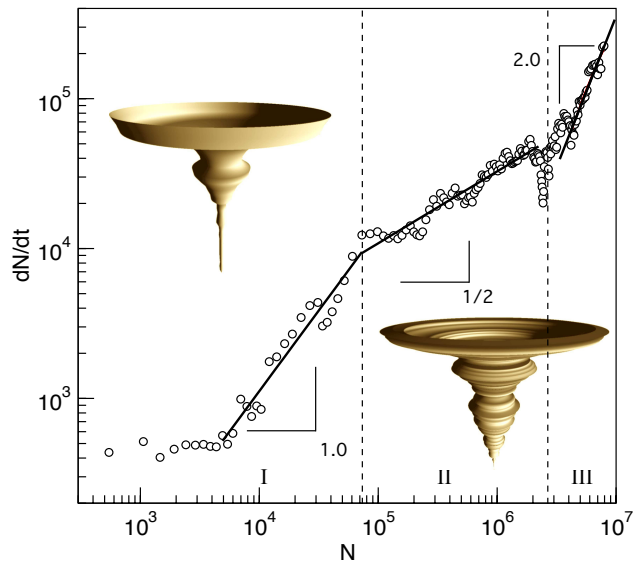}
\caption{Different rates of patent growth are observed in the record of USA patents. 
Here we plot the rate of patent numbers growth $dN/dt$ as a function of $N$. Here we estimate the 
exponents associated to the scaling $dN/dt \sim N^{\beta}$ for three different domains 
that appear to characterise this data set (here indicated as I, II and III). The three exponents $\beta=1,1/2,3/2$ 
indicate different kinetic phases of innovation. The two spindle diagrams (insets) correspond to the first (I) 
interval (left) and the whole time series (right)}
\end{center}
\end{figure}

How is this space expanded? We will assume that every element $\pi_j \in \bf \pi$ 
has the same potential to attach to other existing elements. We can consider a more complex kinetic equation, namely:
\begin{equation}
{dN \over dt}  = \mu(t) N^{1+z}
\end{equation}
where $\mu(t)$ is the attachment rate of inventions at a given time. The parameter 
$0 \le z \le 1$ weights the departure from the linear scenario. For $z=0$ the above yields
to exponential growth. If a pure (pairwise) combination scenario were at work, 
we would see $z=1$, since any pair of inventions is likely to interact.  By solving equation (1) we obtain 
\begin{equation}
N(t) = \left [  
N^{-z}(0)  - z  \int_0^t \mu(\tau) d \tau   \right ]^{-{1 \over z}}
\end{equation}
For the simplest scenario where $\mu$ can be considered constant, i. e. 
$\mu = \langle \mu(t) \rangle =  ( \int_0^t \mu(\tau) d \tau) / t$, 
 we can write the previous equation as follows:
 \begin{equation}
N(t) =  (z \mu)^{-1/z} \left ( t_s - t   \right )^{-1/z} 
\end{equation}
where $t_s$ is a finite time given by $t_s=1/(z\mu N_0^z)$ with $N_0=N(0)$.  
A very interesting feature of this solution is the presence of a {\em singularity}: as 
we approach $t_s$, a divergence occurs in $N$. 
For the pure combination solution with $z=1$ we would observe a growth curve following: 
 \begin{equation}
N(t) =  {1 \over \mu} \left ( {1 \over t_s - t }  \right ) 
\end{equation}
which provides a prediction of how invention numbers will increase under a neutral model 
where all inventions (patents) are equally likely to combine. 
 
 In order to fit the last part ($t>1980$) of the USPTO data set (part III of the time series in figure 2) 
 with the hyperbolic growth prediction and estimate the parameters involved in defining it, we will make 
a logarithmic transformation of the previous general equation as [25]:
  \begin{equation}
\ln N(t) = \ln \Gamma + \eta \ln ( t_s - t )
\end{equation}
 where $\Gamma=(z \mu)^{-1/z}$ and $\eta=-1/z$. Using this transformation, 
 we can estimate the set of parameters associated to the 
 growth process. Following [25], we fixed the critical time $t_s$ to a given year value 
 and fitted the remaining two (free) parameters. By using different $t_s$ values, it is possible 
 to determine the best fitting. For our data set (figure 1, inset), we obtain a critical 
 time $t_s=2027$ and 
 $\eta = 0.87 \pm 0.09$. This gives $z \approx 1.14$ and it would thus approach the expected $\dot{N} \sim N^2$ 
 hyperbolic law\footnote{However deciding what part of the time series to take for the interpolation is far from simple. 
 Several economic, technological and even patent-office related factors influence some of the rapid changes 
 found in the USPTO time series.}. 

 Is this a general result? An interesting 
 observation comes from plotting the rate of patent generation ($dN/dt)$ against the 
 total number of patents $N$ at a given time.  This is displayed in figure 2, where the USPTO 
 data reveal three domains of scaling. These correspond to (I) the industrial revolution 
 (II) the historical period between the end of the IR and the 1990s and (III) the interval 
 $(1990-2010)$. The last part (III) is associated with the aftermath of the 
 information technology revolution associated to the rise of Internet. 
 
 The three domains seem to fit to three 
 different types of kinetics, i e. we would have, following the previous scheme, three 
 different dynamical laws, with $dN/dt=\mu N^{\beta}$ characterised by exponents $\beta_I \sim 1.0, 
 \beta_{II} \sim 0.5$ and $ \beta_{III} \sim 2.0$. For the later domain, we used the last part of the time series from 1980 to the end of the time series. The reason for this choice is the presence of a drop in the time series of patent production 
 that took place due to new management policies by USPTO (12).  
 The last part seems consistent with the minimal hyperbolic model, whereas the first 
 two involve smaller exponents. 
 
 Of note, we can see that the expected linear law associated 
 to the patent dynamics within the IR implies that technological change would lead to exponential 
 growth, whereas the second domain (which includes two World Wars and the great depression) 
 suggests a sublinear dynamical growth equation. Two questions emerge 
 from this seemingly diverse phases. One is whether or not a singularity should be expected and the second 
 the meaning of the scaling exponent $\beta$ . As will be shown below, a generalised model taking into 
 account two key features of the innovation process provide answers to the previous questions.

\section{Generalized models}

If our assumptions were correct, A critical horizon is thus obtained, which indicates that a singularity appears to exist around year 2027. 
Such prediction is performed by using the last part of the time series. However, accelerated patterns 
of innovation growth can be seen at different time intervals, showing different exponents. How can we 
explain these differences? When are singularities expected to occur under these different regimes?
 
Several simplifications have been made in the model above. One is that a limited number of 
previous innovations are combined among them to obtain a new one. Another is that we 
assumed by default that all innovations can contribute to future technologies, when actually 
many will become obsolete and get replaced. Some type of ageing factor needs to be 
considered. Such ageing has been found to be present in different types of growing networks [26-28]
including different forms collaboration among scientists and links among innovations [17]  and will be also studied here.

One way of including multiple innovations is to consider the average number $k$ of 
inventions that are used to obtain new ones. On the other hand, we need to define the 
way two elements might interact. This can be done by considering a generalised integral 
equation: 
  \begin{equation}
\frac{dN}{dt}=\int_{0}^{N}  \buildrel k \over \ldots \int_{0}^{N} \Gamma(\tau_1, ..., \tau_k) d\tau_1... d\tau_k
\end{equation}
 Here the kernel $ \Gamma(\tau_1, ..., \tau_k)$ defines the likelihood that $k$ different patents interact in order to give a new invention. This general expression contains the
 hyperbolic scenario introduced above as one special case when all elements 
 can equally interact and thus $\Gamma=\mu$. To see this, notice that we have now
 \begin{eqnarray}
\frac{dN}{dt}=\int_{0}^{N}  \int_{0}^{N} \mu d\tau_1d\tau_2  \nonumber \\
= \mu \left ( \int_{0}^{N}  d\tau_1 \right ) \left (\int_{0}^{N}  d\tau_2 \right ) = \mu N^2
\end{eqnarray}
From now on we will assume that the kernel can be factorized: all inventions can interact in similar ways and thus 
\begin{equation}
\Gamma(\tau_1, ..., \tau_k) = \prod_{l=1}^k \Gamma(\tau_l)
\end{equation} 
In that case, the previous equation reads now: 
\begin{equation}
\frac{dN}{dt}=\int_{0}^{N}  \buildrel k \over \ldots \int_{0}^{N}  \prod_{l=1}^k \Gamma(\tau_l) d\tau_1... d\tau_k
\end{equation}
and thus our general model to be explored below reads now: 
\begin{equation}
\frac{dN}{dt}= \prod_{l=1}^k  \left [ \int_{0}^{N}  \Gamma(\tau_l) d\tau_l \right ]
\end{equation}

\begin{figure}[htbp]
\begin{center}
\includegraphics[width=0.4 \textwidth]{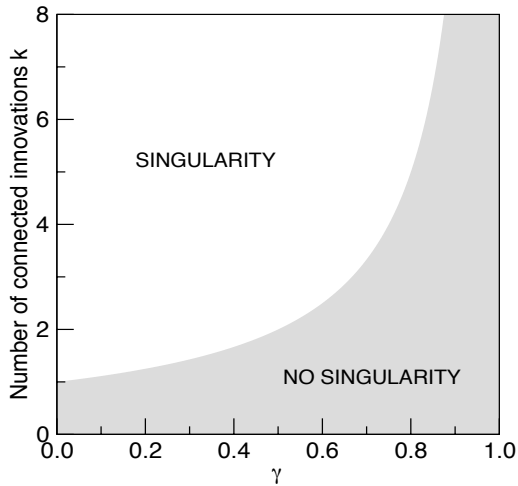}
\caption{Two phases predicted by the generalised model of technological evolution 
with power law ageing. The white area correspond to all parameter combinations allowing 
a singularity to emerge through a hyperbolic growth.}
\end{center}
\end{figure}

\subsection{Power law aging}

This integral equation contains the number of innovations required to further expand the technological space. 
Now we need to introduce how ageing affects the range of interactions. One choice is a power law Kernel, 
namely 
\begin{equation}
\Gamma(\tau_l) \sim \mu^{1/k} (N - \tau_l )^{-\gamma}
\end{equation} 
 where the scaling exponent $\gamma \ge 0$ gives a measure of how fast previous innovations 
 become obsolete and are not incorporated. This Kernel has been used in different contexts, including 
 in the analysis of  collaborations among researchers, which is a closely related problem (15). In this case, 
 the general model is written as
 \begin{equation}
\frac{dN}{dt}= \prod_{l=1}^k  \left [ \int_{0}^{N}  \mu^{1/k} (N - \tau_l )^{-\gamma} d\tau_l \right ]
\end{equation}
 If we assume equivalence between all the components of our system, all kernels being equal we 
 obtain here: 
  \begin{eqnarray}
{dN \over dt}= \left [ \int_{0}^{N(t)} \mu^{1/k} (N - \tau )^{-\gamma} d\tau \right ]^k \nonumber \\
= {\mu \over (1-\gamma)^k} N^{(1-\gamma)k}
\end{eqnarray}

\begin{figure*}
\begin{center}
\includegraphics[width=0.99 \textwidth]{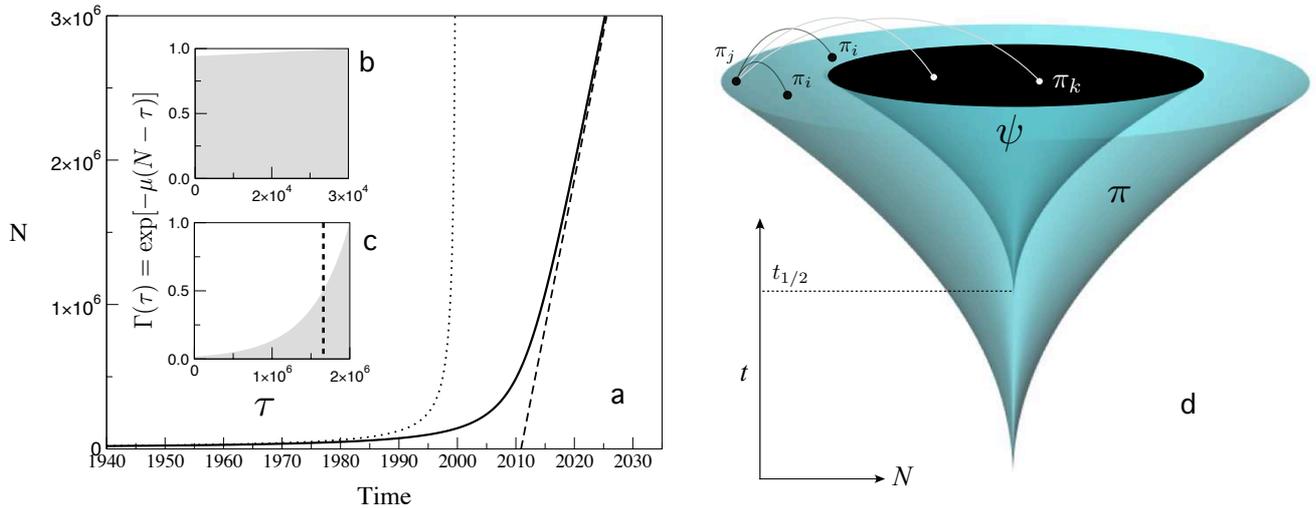}
\caption{Transient hyperbolic growth and blackholes in combination models with limited memory. 
In (a) we display our predicted growth curve $N(t)$ and two approximations 
considering short time (dotted line) and long term (dashed line) scales. The effective kernels 
for these two scales are displayed in the inset plots (b) and (c). The maximum value displayed in the $\tau$ axis of insets (b) and (c) corresponds to $N(t)$ at times $t=1963$ and $t=2020$, respectively. The characteristic scale of ageing imposed 
by the kernel implies that there is a time horizon beyond which no connections among inventions 
can be made. This illustrated schematically in (d) where we shown the spindle diagram of 
the whole system $\pi$ along with a subset $\psi(t)$, first appearing at the characteristic time $t_{1/2}$ where the probability of citing the oldest invention $\pi_1$ is half the maximum. All inventions 
within this "black hole" will be disconnected from the rest. In the present (top large circle) only 
new inventions (filled circles) occupying the outer part of the circle can connect among them whereas 
they cannot link (light lines) with those in the black hole (open circles). The parameter values 
used in (a)-(c) correspond to $\mu =2 \times 10^5$, $t_0 = 1750$, $N_0 = 5000$, and $\gamma=2 \times 10^{-6}$. }
\end{center}
\end{figure*}

 By solving this equation, we can show that the solution reads:
 \begin{equation}
N(t) = \left [ C_0 + \eta(k,n)t \right ]^{1/(1-(1-\gamma)k)}
\end{equation} 
The constants are defined by $C_0=N_0^{1-(1-\gamma)k}$ and
 \begin{equation}
\eta(k,n) =  {1-(1-\gamma)k \over (1-\gamma)^k}
\end{equation} 
respectively.This equation will be consistent with a singularity provided that the scaling exponent is negative. 
This leads to a critical condition:
 \begin{equation}
k > k_c = {1 \over 1 - \gamma}
\end{equation} 
The phase diagram predicted by this critical boundary is shown in figure 3, where we plot 
$k_c(\gamma)$. The two domains showing or lacking a singularity are separate by this curve. 
As we can see, singularities are expected even for $\gamma=0$ provided that $k>1$. Similarly, 
when $k=2$ we have the standard pairwise reaction scheme described above

\subsection{Exponential aging}

The power law kernel introduces a long tail and thus long-memory effects. 
Although links with old inventions are much rare, they can be established and 
thus a contribution will always be expected. What is the impact of using a different 
type of interaction Kernel involving a more rapid decay that forbids new inventions 
to "connect" with very old ones?. This can be modelled with an exponential decay of the form $e^{-\gamma (N-\alpha )}$.
As defined, the smaller the value of $\gamma$, the longer the age (namely, $N-\alpha $) that patents can reach while being
still able to generate new inventions. Indeed, in the limit $\gamma \rightarrow
0$ all the inventions would equally contribute (no matter their age),
recovering again the simple hyperbolic (pairwise) scenario analyzed in section 2.     

In order to illustrate the impact of this limited memory, let us consider again the pairwise 
($k=2$) scenario. The generation of new patents is now given by  
\begin{eqnarray}
\frac{dN}{dt}= \mu \prod_{l=1}^2  \left [ \int_{0}^{N}  e^{-\gamma (N-\tau_l )} d\tau_l \right ] \nonumber \\
= \mu^{\prime }\left[ e^{-\gamma N}-1\right] ^{2}
\end{eqnarray}
where $\mu^{\prime }=\mu/\gamma ^{2}$. By solving this equation we obtain an implicit 
form: 
\begin{eqnarray}
\left(N-N_{0}\right) +\frac{1}{\gamma g_{\gamma}(N)}-\frac{1}{%
\gamma g_{\gamma}(N_0) }+ \nonumber \\
+ \frac{1}{\gamma }\ln \left[ \frac{
g_{\gamma}(N)}{g_{\gamma}(N_0)}\right] =\mu^{\prime }\left( t-t_{0}\right) .
\label{implicit}
\end{eqnarray}
where we used the notation $g_{\gamma}(x) \equiv e^{-\gamma x}-1$. This equation can be 
numerically solved and the result shown in figure 4a for a given set of parameters that was chosen 
to provide a similar range of time and $N$ values than the original series. We can appreciate from this 
diagram that there is a delayed growth phase at the beginning followed by an apparently linear 
growth in late stages. In other words, the dynamics has no singularity. Is that the case? 
Although solving the general problem can be extremely cumbersome, we can deal
with some approximations that can be applied at different stages of the system. 

The first considers the case $\gamma <<1$ and the initial phase of the expansion. If $N$ 
is not large then $\gamma N<<1$ and we can use the approximation $e^{-\gamma N}\simeq 1-\gamma N$. 
on Eq. (\ref{implicit}) leading to:
\begin{eqnarray}
N-N_{0} -\frac{1}{\gamma ^{2}}\left [ {1 \over N} - {1 \over N_0} \right ]+ \nonumber \\
+ \frac{1}{\gamma }\ln \left[ \frac{N}{N_{0}}\right] =\mu^{\prime }\left( t-t_{0}\right). 
\label{initial}
\end{eqnarray}

Given the condition $\gamma << 1$, the LHS of the previous equation is governed by the quadratic terms of $\gamma$. Hence, neglecting the first and last terms on the LHS, we obtain the following approximate solution for the evolution of $N$ at initial stages of technological evolution:
\begin{equation}
N(t)=\frac{1}{\gamma ^{2} \mu^{\prime }\left( t_{s}-t\right) }
\label{initial2}
\end{equation}
where $t_{s}=1/(\mu^{\prime } \gamma ^{2}N_{0})+t_{0}$. The previous equation 
predicts a singularity at some time in the future although such singularity is in conflict 
with our approximation and the hyperbolic growth is only a transient phenomenon.

Let us know focus on the long-term ($t\rightarrow \infty $) dynamics of the
system. In this case it is sensible to assume a very large number of
existing patents, and hence $e^{-\gamma N}\rightarrow 0$. In this case, we can
rewritte Eq. (\ref{implicit}) as: 
\begin{equation}
\left( N-N_{0}\right) -\frac{1}{\gamma }+\frac{1}{\gamma ^{2}N_{0}}-\frac{1}{\gamma }%
\ln \left[ \gamma N_{0}\right] =\mu^{\prime }\left( t-t_{0}\right) ,
\label{longterm}
\end{equation}
where we have also used again $e^{-\gamma N_{0}}\simeq 1-\gamma N_{0}$. 
Thus, from Eq. (\ref{longterm}) it is
straightforward to obtain the long-term solution for the dynamics of $N$:
\begin{equation}
N(t)=\mu^{\prime }\left( t-t_{0}\right) +\varepsilon   \label{longterm2}
\end{equation}
with $\varepsilon =N_{0}+1/\gamma-1/(\gamma ^{2}N_{0})+\ln \left[ \gamma N_{0}\right]/\gamma$.  

Equation (\ref{longterm2}) reveals that $N$ exhibits a linear growth
dynamics when large values of $t$ are considered. Note that this long-term
dynamics is notably different from the hyperbolic dynamics predicted for
initial stages of evolution (see the explanation above). In figure 4 we show the 
fits of these approximations to the exact solution (obtained numerically). As we can see, 
the analytic results confirm that the initial hyperbolic trend (dotted curve)is eventually replaced 
by a slowdown characterised by a linear process (dashed line) with no technological singularity associated. 

The previous results can be generalised to the $k$-diversity scenario, where the new 
equation reads
\begin{equation}
\frac{dN}{dt}=(-1)^{k+1} \mu \prod_{l=1}^{k}\left[
\int_{0}^{N}e^{-\gamma (N-\tau _{1})}d\tau _{1}\right] 
\end{equation}
where the term $(-1)^{k+1}$ is required in order to avoid the unphysical situation in
which the number of patents decreases when combining an odd number of
previous inventions (note that a negative integration constant is obtained
when integrating $e^{-x}$).

This general model leads to: 
\begin{equation}
\int_{N_{0}}^{N}\frac{d\widetilde{N}}{\left[ e^{-\gamma \widetilde{N}}-1\right]
^{k}}=\mu^{\prime }\int_{t_{0}}^{t}d\widetilde{t}
\end{equation}
which, at initial stages reads:
\begin{equation}
\int_{N_{0}}^{N}\frac{d\widetilde{N}}{\left( \gamma \widetilde{N}\right) ^{k}}= \mu^{\prime }\int_{t_{0}}^{t}d\widetilde{t},
\end{equation}
and gives hyperbolic dynamics, whereas in the long-term dynamics we have now:
\begin{equation}
\int_{N_{0}}^{N}d\widetilde{N}= \mu^{\prime }\int_{t_{0}}^{t}d\widetilde{t},
\end{equation}
again leading to linear dynamics.
 
 An intuitive explanation for the change from early hyperbolic to late linear dynamics is provided 
 in the insets of figure 4b-c. Here we show the kernel associated to early (fig 4b) 
 and late (fig 4c) times, and thus smaller and larger numbers of innovations. Although the 
 area covered by $\Gamma$ almost fills the plot when $N \sim O(10^4)$, it becomes smaller with large $N$ values 
 (here $N \sim O(10^6)$). In Fig. 4b, the probability of recombining any existing patent is higher than $90\%$ of the maximum [i.e., $90\%$ of the probability of recombining the newest pattent $N(t)$, with $t=1963$]. However, the probability of combining patents existing at $t=2020$ (Fig. 4c) folds to approximately zero for the oldest patents. Then, we can arbitrarily define a patent number $\pi_{h} (t)$ that delimits the frontier between up-to-date patents and obsolete patents (which will hardly ever been recombined again). Specifically, we consider $\pi_{h} (t)$ to be the patent number for which the probability of recombination is half the maximum (as indicated by the dashed line in Fig. 4c). Thus, the effect of ageing (or loss of memory) dominates in the long term (Figs. 4a and c) and the effective rate of innovation 
 become linear. Such slowdown prevents the system from approaching a divergent dynamics. 
 
 These results can be graphically interpreted as shown in figure 4d. Here we use again the spindle diagram 
 showing how the universe (or space) of innovations experiences an accelerated growth at early stages of development. Novel 
 patents such as $\pi_j$ will be distributed over the outer parts of the patent space (a circle at each time step) 
 and connect with others such as $\pi_i$. 
 After a critical time $t_{1/2}$, some inventions start to become obsolete or forgotten. From this time on, the expansion speed stabilises, and both the universe of inventions and the "blackhole" $\psi(t)$ at its center (which represents the area of obsolete technology) grow at the same constant speed. Inventions within the blackhole (such as $\pi_k$ in figure 4d) cannot be used and thus no 
 information about them can cross the obsolescence frontier. Our technological memory establishes the distance between these two boundaries in the innovations space, and this distance determines the number of up-to-date inventions, which in turn determine the expansion rate of the innovation space.

\section{Discussion}

The nature and tempo of innovation is a difficult and timely topic. It has been 
the focus of attention from evolutionary biologists, economists and physicists alike. 
Inventors get inspiration from previous, existing designs, while they push forward 
the boundaries of innovation. In searching for a theory of technological change, 
the combinatorial nature of technological innovation seems to be an essential 
component of human creativity. By combining previous designs into novel inventions, 
there is a potential for an explosion of novelties, which could eventually 
move towards a singularity. How can we test such possibility? 
Patent citations are a privileged window into such process, since they  
provide a first approximation to both the growth of inventions and their 
interactions over time. The accelerated pattern of patent growth suggests that 
a super linear process of innovation is taking place and available evidence indicates that 
this is at least partially associated to combinatorial processes (18). 

In this paper we have explored a simple class of models that include both the richness of combinations 
and how rapidly the relevance of previous inventions fades with time. These two features can be 
seen as two opposing forces: the diversity of potential previous inventions to be combined 
powers combinatorial design, while the obsolescence of the same inventions 
makes them less likely to contribute to combinations. Our goal was not 
as much as to fit data than understand the basic scenarios where singularities might 
emerge when both features are included. 

We have shown that long-memory kernels 
permit the presence of singularities under some conditions, while kernels involving 
a characteristic time scale of ageing forbid divergences to occur. The first 
class predicts two different phases, which reminds us of a picture of 
innovation defining a phase transition between sub-critical and super-critical phases (29). 
The second provides a plausible explanation of why singularities might fail to be observed, 
while the transient dynamics of innovation appears hyperbolic.  
Further investigations should analyse other temporal trends (including the patterns of fluctuations) associated 
to these class of models and a more details analysis of available time series.  Existing models of 
evolution of innovations (30,31) can provide very useful tests to the ideas outlined here.

%%%%%%%%%%%%%%%%%%%%%%%%%%%%%%%%%%%%%%%%%%%%%%%%%%%%

\vspace{0.5 cm}

\noindent
{\bf Acknowledgments}

\vspace{0.5 cm}

RS and SV thank Marti Rosas-Casals, Stuart Kauffman, Doyne Farmer and Niles Eldredge 
for past discussions on cultural evolution. We also thank the members of the CSL for many stimulating 
discussions. This paper is dedicated to the defenders of the last barricade before the Eglise Sant-Merri. 
This work was supported  by the Fundacion Botin and by the Santa Fe Institute, 
where most of this work was done.

\vspace{0.5 cm}

%%%%%%%%%%%%%%%%%%%%%%%%%%%%%%%%%%%%%%%%%%%%

\bibliography{pre} % Produces the bibliography via BibTeX.

%%%%%%%%%%%%%%%%%%%%%%%%%%%%%%%%%%%%%%%%%%%%

\section{References}

\begin{enumerate}

\item
Basalla, (1989) {\em The evolution of technology}. Cambridge U. Press.

\item
Arthur, B. (2009) {\em The nature of technology. What it is and how it evolves}. Free Press.

\item
Kelly, K (2010) {\em What technology wants}. Viking.

\item
Johnson S (2010) {\em Where good ideas come from: a natural history of innovation}. Riverhead Press.

\item
Sahal, D. (1981). {\em Patterns of technological innovation}. Addison-Wesley.

\item
Jacob, F. (1977). Evolution as tinkering. Science 196, 1161-1166.

\item
Sol\'e, R.V., Ferrer, R., Montoya, J. M. and Valverde, S. (2002). 
Selection, Tinkering, and  Emergence in Complex Networks. Complexity 8, 20-33.

\item
Schuster, P. (2006) Untamable curiosity, innovation, discovery and bricolage. Complexity 11, 9-11.

\item
T\"emkin, I. and Eldredge, N. (2007). Phylogenetics and Material Cultural Evolution. Curr. Antrop. 48, 146-153.

\item
Eldredge, N. (2011) Paleontology and cornets: Thoughts of material cultural evolution. Evo. Edu. Outreach 4, 364-373.

\item
McNerney, J., Farmer, J. D., Redner, S. and Trancik, J. E. (2011)  Role of design complexity in technology improvement, 
Proc. Natl. Acad. Sci USA 108(22), 9008-9013.

\item
Sol\'e, R.V., Valverde, S., Rosas-Casals, M., Kauffman, S.A., Farmer, J.D., and Eldredge, N.(2013) 
The evolutionary ecology of technological innovations, Complexity 18(4), 15-27.

\item
Kurzweil R (2005) {\em The singularity is near}. New York : Viking.

\item
Nagy, B., Farmer, J.D., Trancik, J. E. and Gonzales, J.P. (2011)  
Superexponential Long-Term Trends in Information Technology. J. Tech. Forecast. Soc. Change 73, 1061-1083.

\item
Nagy, B., Farmer, J.D., Bui, Q.M. and Trancik, J. E. (2013) 
Statistical basis for predicting technological progress. PLOS ONE 8: e52669. 

\item
Jaffe, AB and Trajtenberg, M (2003) {\em Patents, Citations and Innovations}. MIT Press.

\item
Valverde, S, Sol\'e RV, Bedau M. and Packard, N (2007) 
Topology and evolution of technology innovation networks. Phys. Rev. E 76, 032767. 

\item
Youn, H., Bettencourt, L.M.A., Strumsky, D. and Lobo, J (2014) Invention as a combinatorial process: evidence from U. S. patents. arXiv:1406.2938 [physics.soc-ph]. 

\item
Eigen, M., and Schuster, P. (1978) Part A: Emergence of the Hypercycle,  Naturwissenschaften 65, 7�41.
 
\item
Sol\' e, R.V.  (2011) {\em Phase Transitions}. Princeton U. Press.

\item
Hanel, R, Kauffman SA and Thurner S (2005) Phase transitions in random catalytic networks. Phys. Rev. E 72, 036117.

\item
Hanel, R, Kauffman SA and Thurner S (2007) Towards a physics of evolution: critical diversity dynamics at the edges of collapse and bursts of diversification. Phys. Rev. E 76, 036110.

\item
von Foerster, H., Mora, P. M., and Amiot, L. W. (1960)  Doomsday: Friday, November 13, AD 2026, Science 132, 1291�1295.

\item
Cohen, J. E. (2003) Human Population: The Next Half Century, Science 302 (5648), 1172�1175.

\item
Johansen, A., and Sornette, D. (2001) Finite-time singularity in the dynamics of 
the world population and economic indices, Physica A 294 (3-4), 465-502. 

\item
Shao, Z.G., Zou, X.W., Tan, Z. J. and Jin, Z. Z. (2006) Growing networks with mixed attachment mechanisms. 
J. Phys. A 39, 2035-2041. 

\item
S. N. Dorogovtsev and J. F. F. Mendes, (2000) Evolution of networks with aging of sites. Phys. Rev. E 62, 1842-1845. 

\item
Borner, K.,  Maru, J.T. Goldstone, R.L.. (2004) The simultaneous evolution of author and paper networks. 
Proc. Natl. Acad. Sci. U.S.A. 101, 5266-5273.

\item
Kauffman, S. A. (1995) {\em At home in the universe}. Oxford U. Press. New York. 

\item
Arthur, B. and Polak,W. (2006) 
The evolution of technology within a simple computer model. Complexity 11, 23-31.

\item
Thurner, S., Klimek, P. and Hanel, R (2010) Schumpeterian economic dynamics as a quantifiable model of evolution. 
New J. Phys. 12, 075029.

\end{enumerate}
\end{document}